\documentclass[twoside,leqno,twocolumn]{article}

\usepackage[letterpaper]{geometry}

\usepackage{ltexpprt}
\usepackage{hyperref}
\usepackage{graphicx}
\usepackage{xcolor}
\usepackage{algorithm}
\usepackage{algpseudocode}
\usepackage{amsmath,amssymb,amsfonts}
\usepackage{subcaption}
\usepackage{siunitx}

\newcommand{\email}[1]{\protect\href{mailto:#1}{#1}}

\newenvironment{@abssec}[1]{%
     \if@twocolumn
       \section*{#1}%
     \else
       \vspace{.05in}\footnotesize
       \parindent .2in
         {\upshape\bfseries #1. }\ignorespaces 
     \fi}
     {\if@twocolumn\else\par\vspace{.1in}\fi}

\newcommand\keywordsname{Key words}
\newenvironment{keywords}{\begin{@abssec}{\keywordsname}}{\end{@abssec}}

\DeclareMathOperator*{\argmin}{arg\,min}

\begin{document}

\newcommand\relatedversion{}

\title{ProGReST: Prototypical Graph Regression Soft Trees for Molecular Property Prediction }
\author{Dawid Rymarczyk\thanks{Faculty of Mathematics and Computer Science, Jagiellonian University, Krakow, Poland. (\email{dawid.rymarczyk@student.uj.edu.pl}, \email{daniel.dobrowolski@student.uj.edu.pl}, \email{tomasz.danel@doctoral.uj.edu.pl})}\ \thanks{Ardigen SA, Krakow, Poland.}
\and Daniel Dobrowolski\footnotemark[1] \and Tomasz Danel\footnotemark[1]}

\date{}

\maketitle

\fancyfoot[R]{\scriptsize{Copyright \textcopyright\ 2023 by SIAM\\
Unauthorized reproduction of this article is prohibited}}

\begin{abstract} \small\baselineskip=9pt

In this work, we propose the novel Prototypical Graph Regression Self-explainable Trees (ProGReST) model, which combines prototype learning, soft decision trees, and Graph Neural Networks. In contrast to other works, our model can be used to address various challenging tasks, including compound property prediction. In ProGReST, the rationale is obtained along with prediction due to the model's built-in interpretability.  Additionally, we introduce a new graph prototype projection to accelerate model training. Finally, we evaluate PRoGReST on a wide range of chemical datasets for molecular property prediction and perform in-depth analysis with chemical experts to evaluate obtained interpretations. Our method achieves competitive results against state-of-the-art methods.
\end{abstract}

\begin{keywords}
Drug design; Graph Neural Networks; Interpretability; Deep Learning 
\end{keywords}

\section{Introduction}

In chemistry, the accurate and rapid examination of the compounds is often the key to a successful drug discovery. Searching through millions of compounds, synthesizing them, and evaluating their properties consumes astounding amounts of money and does not guarantee any success at the end of the discovery process. That is why currently \textit{in silico} molecular property prediction is indispensable in modern drug discovery, material design, synthesis planning, etc. Computer methods can accelerate compound screening and mitigate the risk of selecting futile compounds for the \textit{in vitro} examination.

Recent advancements in deep learning, especially in Graph Neural Networks (GNNs), raised the usability of \textit{in vitro} cheminformatics tools to the next level~\cite{david2020molecular}. Tasks such as molecular property prediction, detection of active small molecules, hit identification, and optimization can be accelerated with models such as  Molecule Attention Transformer (MAT)~\cite{maziarka2020molecule}, DeepGLSTM~\cite{mukherjee2022deepglstm}, and Junction Tree Variational Autoencoder (JT-VAE)~\cite{jin2018junction}. Despite the early adoption of artificial intelligence (AI) methods in the drug design process, the initial results are encouraging~\cite{mak2022success}. Unfortunately, most AI methods do not offer insight into the reasoning behind the decision process. 

\begin{figure}
    \centering
    \includegraphics[width=0.4\textwidth]{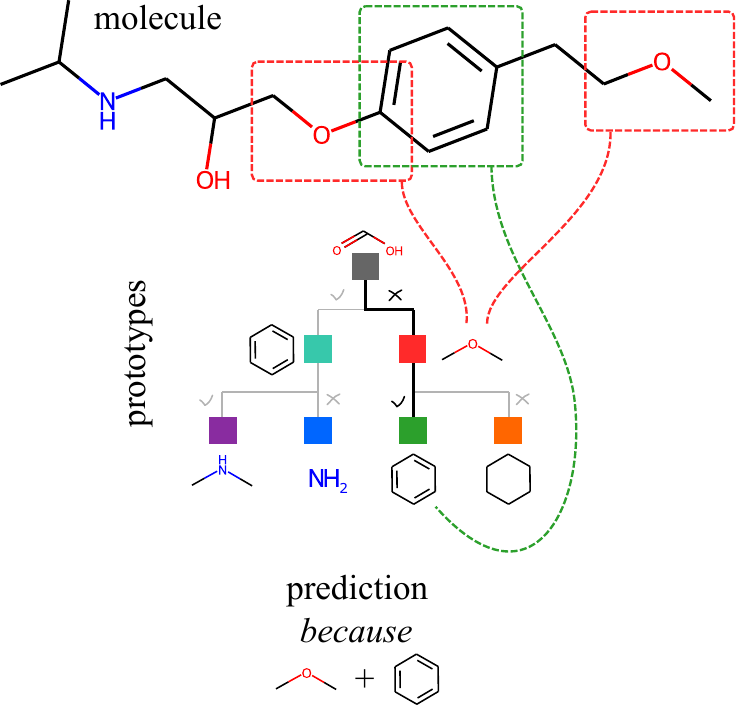}
    \caption{Overview of the ProGReST approach. Molecular substructures are matched against the trained prototypical parts, and the prediction is based on the combination of these features.}
    \label{fig:first}
\end{figure}

Due to the complexity of biological systems and drug design processes, insights into the knowledge gathered by the deep learning model are highly sought. Even if the model fails to achieve its goals, the explainability component can hint at the medicinal chemist, e.g. by showing a mechanistic interpretation of the drug action~\cite{jimenez2020drug}. Most of the current eXplainable Artificial Intelligence (XAI) approaches are post-hoc methods and are applied to already trained models~\cite{ying2019gnnexplainer}. However, the reliability of those methods is questionable~\cite{rudin2019stop}. It assumes that the second model is built to explain an existing trained model. It may result in an unnecessarily increased bias in the explanations, which come from the trained model and the post-hoc model. That is why self-interpretable models are being developed, such as self-explainable neural networks (SENN)~\cite{NEURIPS2018_3e9f0fc9} and Prototype Graph Neural Network (ProtGNN)~\cite{zhang2021protgnn}. Only the latter can be applied to the graph prediction problem.

However, ProtGNN is designed for classification problems only since it requires a fixed assignment of prototypes to the classes. While for a regression problem, the model predicts a single label making such an assignment impossible. To overcome the limited applicability of ProtGNN, we introduce the Prototypical Graph Regression Soft Trees (ProGReST) model that is suitable for a graph regression problem, common in the molecular property prediction~\cite{wieder2020compact}. It employs prototypical parts (in the paper, we use the terms "prototypical parts" and "prototypes" interchangeably.)~\cite{chen2019looks} that preserve information about activation patterns and ensure intrinsic interpretability (see Fig.~\ref{fig:first}). Prototypes are derived from the training examples and used to explain the model's decision. To build a model with prototypes, we use Soft Neural Trees~\cite{frosst2017distilling}. 

Hence the regression task is more challenging than the classification, it also requires more training epochs for a model to converge. And, prototypical-part-based methods use projection operation periodically~\cite{chen2019looks,zhang2021protgnn} to enforce the closeness of prototypes to the training data. In ProtGNN, projection is based on an MCTS algorithm that requires lots of computational time to find meaningful prototypes. In ProGReST, we propose proxy projection to reduce the training time and perform MCTS-based at the end to ensure the full interpretability of the derived prototypes. 

The ProGReST achieves state-of-the-art results on five cheminformatics datasets for molecular property prediction and provides intuitive explanations of its prediction in the form of a tree. Also, we confronted the findings of the ProGReST with chemists to validate the usability of our model.

Our contributions can be summarized as follows:
\begin{itemize}
    \item we introduce ProGReST, a self-explainable prototype-based model for regression of molecular properties, 
    \item we employ a tree-based model to derive meaningful prototypes,
    \item we define a novel proxy projection function that substantially accelerates the training process.
\end{itemize}

\section{Related Works}

\subsection{Molecular property prediction}
The accurate prediction of molecular properties is critical in chemical modeling. In machine learning, chemical compounds can be described using calculated molecular descriptors, which are computed as a function of the compound structure~\cite{todeschini2008handbook}. Many successful applications of machine learning in drug discovery utilize chemical structures directly by employing molecular fingerprints~\cite{capecchi2020one} or molecular graphs as an input to the model~\cite{gaudelet2021utilizing}.

Currently, molecular graphs are a preferable representation in cheminformatics because they can capture nonlinear structure of the data. In a molecular graph, atoms are represented as nodes, and the chemical bonds are graph edges. Each atom is attributed with atomic features that encode chemical symbols of the atom and other relevant features~\cite{pocha2021comparison}. This graphical representation can be processed by graph neural networks that learn the molecule-level vector representation of the compound and use it for property prediction. Graph neural networks usually implement the message passing scheme~\cite{gilmer2017neural}, in which information is passed between nodes along the edges, and the atom features are updated~\cite{zhang2019graph}. However, more recent architectures focus on modeling long-range dependencies between atoms, e.g. by implementing graph transformers~\cite{maziarka2021relative}.

\subsection{Interpretability of deep learning}
Methods explaining deep learning models can be divided into the post-hoc and interpretable~\cite{rudin2019stop}. The first one creates explainer that reveals the reasoning process of a black box model. Post-hoc methods include: a saliency map~\cite{baldassarre2019explainability} that highlights crucial input parts. Another one is Concept Activation Vectors (CAV), that uses concepts to explain the neural network predictions~\cite{kim2018interpretability}. Other methods analyze the output of the model on the perturbation of the input~\cite{ribeiro2016should} or determine contribution of a given feature to a prediction~\cite{yuan2021explainability}. Implementation of post hoc methods is straightforward since there is no intervention into its architecture. However, they can produce biased and unreliable explanations~\cite{NEURIPS2018_294a8ed2}. That is why more focus is recently on designing self-explainable models~\cite{NEURIPS2018_3e9f0fc9} to make the decision process directly visible. Recently, a widely used self-explainable model introduced in~\cite{chen2019looks} (ProtoPNet) has a hidden layer of prototypes representing the activation patterns.

Many of the works extended the ProtoPNet, such as TesNet~\cite{wang2021interpretable} employing Grassman manifold to find prototypes. Also, methods like ProtoPShare~\cite{rymarczyk2021protopshare}, ProtoPool~\cite{rymarczyk2021interpretable} and ProtoTree~\cite{nauta2021neural} reduce the number of used prototypes. Lastly, those solutions are widely adopted in various fields such as medical imaging~\cite{kim2021xprotonet} and graph classification~\cite{zhang2021protgnn}. Yet, none of these do not consider regression. 

\begin{figure*}[ht]
\centerline{\includegraphics[width=\linewidth]{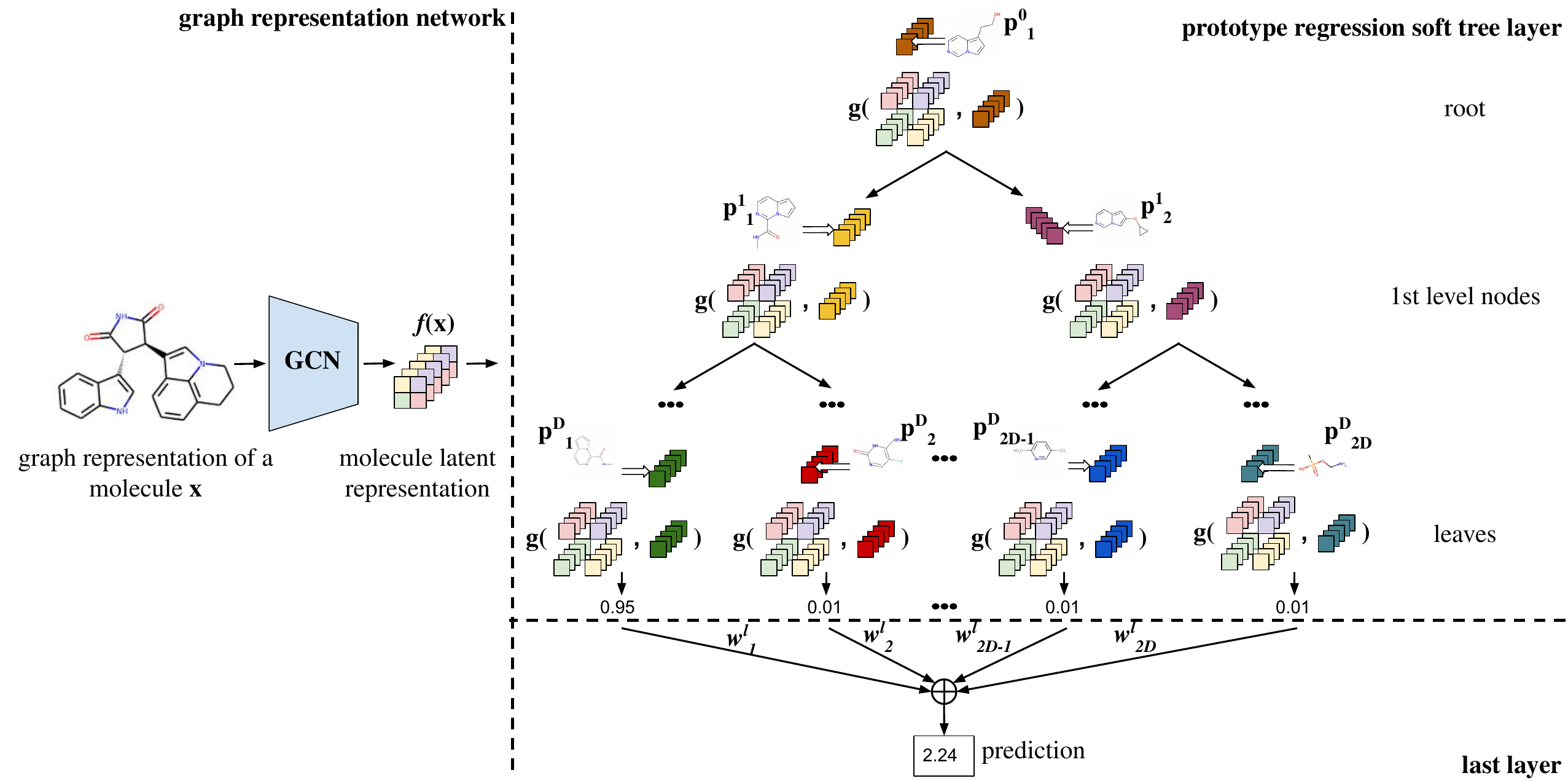}}
\caption{ProGReST architecture. It consists of a graph convolutional neural network (GCN) that generates the latent representation of the molecular graph. Later on in the prototype regression, the soft tree layer computes the similarity of each node (prototypical part) to each latent vector of molecular representation. Then the last layer is used to obtain the prediction.}
\label{fig:arch}
\end{figure*}

\section{ProGReST}

\subsection{Architecture}

The architecture of ProGReST, depicted in Fig.~\ref{fig:arch}, consists of a graph representation network $f$, a prototypical regression soft tree layer $t$ and the last layer $h$. We consider a regression problem with a dataset consisted of $K$ graphs $x_i \in \mathcal{G}$ with corresponding labels $y_i \in \mathcal{Y}\subset{\mathbb{R}}$. Graph $\mathcal{G} \subset{\mathbb{R}}^{\mathcal{N} \times \mathcal{E}}$ contains a set of possible nodes $\mathcal{N}$ and a set of graph edges $\mathcal{E}$. Given an input graph $x_i \in \mathcal{G}$, the model returns its prediction $\hat{y_i}\in\mathcal{Y}$. 

Before a molecule is processed by the model, it is encoded as an array of shape $N \times P$, where $N$ is a number of nodes and $P$ is the size of the vector encoding each node. Then, the graph representation network is used to calculate the input embedding $z=f(x)$ where $z\in \mathbb{R}^{N \times C}$, $C$ is the prototype depth, and $x$ is an input graph. The graph representation network is a graph convolutional network (GCN)~\cite{kipf2016semi} followed by a node-wise convolution with the sigmoid activation at the end, used to map the input features to the prototype space. The additional node-wise layers reduce the dimensionality of the latent representation and facilitate the learning of meaningful prototypical parts.

The prototype regression soft tree layer contains $2^{\mathcal{D}-1}$ prototypes $p_j\in\mathbb{R}^{C}$, where $\mathcal{D}$ is the depth of the tree. The number of prototypes is the same as the number of nodes of the tree because there is only one prototypical part in each node. All nodes have two children and the tree contains $2^{\mathcal{D}}$ leaves. Each leaf $l_i$ calculates only one value $y_{l_i}\in\mathbb{R}$. The prototypes are trainable parameters.

For each input $x \in \mathcal{G}$, ProGReST calculates the prototype activation as the similarity between the prototypical part $p$ and the latent graph representation $z_i\in\mathbb{R}^C$ for each graph node $i$. Then, it calculates the maximum activation as a \textit{presence} of a given prototype in the input graph:

\begin{equation}
\hat{z} = \max_i e^{-||z_{i}-p||_2}
\label{eq:similarity}
\end{equation}

Unlike decision trees with nodes routing to only one child, Soft Decision Trees distribute the signal to both children simultaneously, with the probability adding up to $1$. To assure that in each node there is a single prototypical part with a similarity value from Equation~\ref{eq:similarity} used as a probability of routing to a right node as it is in ~\cite{nauta2021neural}. The  probability of routing to the left node is a complement to $1$ and equals $ 1 - \hat{z}$.

To determine the probability in a leaf $l_k$, we need to traverse through the path $\mathcal{P}_i$ consisted of its parents:

\begin{equation}
\label{eq:path_prob}
l_{k}(x) = \prod_{\hat{n}\in \mathcal{P}_k} \hat{z}_{\hat{n}}(x),
\end{equation}
where $\hat{z}_{\hat{n}}$ is the similarity value in node $\hat{n}$.
Then, the final prediction is made by summing up probabilities from the leaves multiplied by weights $w^l_k$ of the last layer $h$:

\begin{equation}
\hat{y}(x) = \sum_{l_k\in \ell} w^l_k \cdot l_k(x).
\end{equation}

\subsection{Regularizers}

Additional regularizers in the loss function of ProGReST are added to ensure that the prototypes and the soft decision tree are effectively learnt. 

First of all, we want to minimize the chances of the tree routing only to the left side of the tree. The origin of this problem is in the initialization of the model. It is difficult to activate equally prototypes due to their large distance from the prototypical parts. To mitigate this we implement  regularization from~\cite{frosst2017distilling} that encourages each node to use the left and right subtrees equally. The penalty is the cross entropy between distribution $[0.5, 0.5]$ and the actual average distribution [$\alpha_{i}$, $1 - \alpha_{i}$] where $\alpha_{i}$ for node $i$ is given by:
\begin{equation}
\alpha _{i} = \frac {\sum _{x}\Pi_{i}(x)\hat{z}_{i}(x)}{\sum _{x}\Pi_{i}(x)},
\end{equation}
where $\Pi_{i}(x)$ is the path probability from the root to node $i$. Next, we calculate the weighted sum of the cross entropy, weighted by $2^{-d_i}$ between each node and balanced distribution. $d_i$ is the depth of node $i$.

\begin{equation}
L_{p} = -\sum _{i=1}^ {2^{D-1}}2^{-d_{i}}\left[{0.5\log (\alpha _{i}) + 0.5\log (1-\alpha _{i})}\right].
\end{equation}

As in other prototypical-parts-based models, such as~\cite{chen2019looks,rymarczyk2021protopshare,zhang2021protgnn}, the regularization of a latent space is needed to derive meaningful prototypes. For that purpose we adapt the cluster cost from~\cite{kim2021xprotonet} to assure that the prototypes are close to the parts of the training data points with a batch of $B$ examples. 

\begin{equation}
L_{c} = \frac{1}{|B|} \sum_{x \in B} \min_{p_j \in \mathcal{P}} \min_{z \in f(x)} ||z-p_j||_2.
\label{eq:clst_cost}
\end{equation}

The penalty function given by Eq.~\ref{eq:clst_cost} may enforce the prototypical parts to be identical. To prevent that we propose a novel orthogonal loss between parents of each leaf. We penalize the similarity between prototypical parts between parents and not all nodes because some of the prototypes can be reused on a different path.   
For each leaf we have $\mathcal{D}$ parents. The goal here is to have a have different prototype in each of the node on the given path to maximize the capacity of the model. We propose to minimize the Frobenius norm  between the leaf parents.

\begin{equation}
L_{d} = \frac{1}{|\ell|} \sum_{l \in \ell} \sqrt{
\sum_{p_i \in \mathcal{P}_l}  \sum_{j > i, p_j \in \mathcal{P}_l} 
|S_c(p_i, p_j)|^2},
\end{equation}
where $S_c$ is a cosine similarity. To ensure the majority of probability mass in a single leaf, we use a modified mean squared error ($MSE_W$) in the training phase. It is needed to assure ease to understand interpretation focused on a single path to a leaf. Given the label $y$ for input $x$, we define the loss function as follows:

\begin{equation}
MSE_W(y, w^l) = \frac{1}{|\ell|}\sum_{l_k\in \ell} l_k(x) \cdot (y- w^l_k)^2,
\end{equation}
where $w^l_k$ is the weight of a given leaf in the last layer. When the probability in a given node is $1$, the prediction itself is $w^l_k$. So the single path is used to derive the prediction. As a results the final cost function is as follows:

\begin{equation}
L = MSE_W(y, w^l) + \lambda_pL_p + \lambda_c+L_x + \lambda_dL_d,
\label{eq:loss}
\end{equation}
where $\lambda_p$, $\lambda_c$, $\lambda_d$ are hyper-parameters. 

\subsection{Proxy projection}
To assure that prototypes are from the training dataset distribution, we need to use a projection that swaps the prototypical parts with a vector from graph latent representation. In~\cite{zhang2021protgnn}, a projection is based on the MCTS algorithm. However, it is done periodically during the training and is very computationally expensive. That is why we use a proxy projection for periodic prototype assignments while at the end of the training we perform the MTCS-based one. Our proxy projection is much faster than MCTS, but also not-interpretable. However, it can be used as an approximation of the exact one. 

Our proxy projection tries to find the closest vector from latent graph representation to a given prototypical part and replace it. However, such an assignment makes it non-interpretable since we do not know which nodes with what level of importance contributed to a given representation vector. The proxy projection is defined as follows: 

\begin{gather} 
\label{eq:project_patch}
p \gets \argmin _{z\in Z_j} || z-p_j||_2, \\
\nonumber
\text{where }
Z = \{ \hat{z}: \forall{i},  \hat{z} \in f(x_i)\}.
\end{gather}

The MCTS-based projection looks for a subgraph of the input graph, which is the closest to the prototype in the latent space~\cite{zhang2021protgnn}. This method can be easily interpreted because two subgraphs shall similarly activate the same prototype. The MCTS-based projection:
\begin{gather} 
\label{eq:project_mcts}
p_{j} \gets \argmin _{z\in Z_j} || z-p_j||_2, \\
\nonumber
\text{where } Z_j = \{ \hat{z}: \forall{i}, \text{pool}(f(\hat{x})), \hat{x} \in \text{Subgraph}(x_i)\}.
\end{gather}

In addition, we reduce the time needed for MCTS-based projection to find the closest subgraph by limiting the search to a single graph. Firstly, with our proxy projection, we identify the graph with the most similar latent vector to a prototype, and then we apply an MCTS-based search on the found graph to identify significant vertices.

\subsection{Training schema}

Due to the more challenging nature of regression comparing to classification~\cite{zhang2019graph}, we introduce an additional training part called a warmup. During warmup, we firstly cluster the molecules using the K-Means algorithm based on the molecular property values. Then we assign each molecule to a given cluster and we treat it as a new label of a given molecule. Using those pseudo-labels we pre-train full ProGReST using all regularizers with a CrossEntropy Loss $L_{CE}$ using training schema from~\cite{nauta2021neural}.

At the end of the warmup, each leaf weight $w^l$ is translated into scalars using K-Means centroids. For each leaf $l$ we have:
\begin{equation}
\label{eq:translate_leaf}
l = \operatorname{Softmax}(l)\cdot\mathcal{K}_{centroids}.
\end{equation}

Similarily to~\cite{chen2019looks}, warmup allows the model to derive initial prototypical parts that can be reused in a regression task.
After warmup we train ProGReST using loss function from Eq.~\ref{eq:loss}. Starting from a given epoch, we periodically perform a proxy projection Eq.~\ref{eq:project_patch}. After the last projection (MCTS-based one) we only train leaves to refine the model. The detailed algorithm for training can be found in Supplementary Materials. 

\section{Experimental Setup}
To evaluate the proposed model, we used five datasets from PyTDC~\cite{Huang2021tdc} that are dedicated to predicting molecular properties. Depending on the dataset, we followed the fold splitting strategy recommended for a given set (scaffold-based split or random split). Each of the datasets was divided into training, validation, and testing sets with the following proportions $80\%$, $10\%$, and $10\%$ respectively. 

All experiments were implemented in Python 3 and the model is implemented in PyTorch library~\cite{paszke2019pytorch}\footnote{The code is available at \url{https://github.com/gmum/ProGReST}}. As a graph representation generation network $f$, we use graph convolutional network 
\cite{gcn_conv} containing $3$-$5$ layers with $128$-$512$ kernels. We used prototypes of depth $C \in \{64, 128, 256\}$. For each dataset, we performed a grid search of hyperparameters of the model. Among the tested parameters were: tree depth $\mathcal{D} \in [4;7]$, warmup pseudo labels number from $1$ to $2^{\mathcal{D}-1}$ so that each class can be in at least one leaf. The warmup lasts up to $80$ epochs with an early stopping period of $5$ epochs. In the next phase of training, we learned ProGReST for $250$ epochs and again used an early stopping mechanism with a window of $10$ epochs. Weights of the regularizers from the loss functions were: $\lambda_c \in [0.05, 0.8]$, $\lambda_p \in [0.05, 0.3]$ and 
$\lambda_d \in [0.0001, 0.01]$.
For the MCTS-based projection, we limited the number of iterations to $32$. A minimum number of atoms in MCTS-based projection is set to $3$ and a maximum to $12$. Moreover, MCTS can expand to $12$ children. Periodical projection starts at $80^{th}$ epoch and is performed every $20$ epoch. Each model was run 5 times with different seeds. As an optimizer, we use ADAM~\cite{kingma2014adam} with a learning rate $\eta$ different for each node $p_j$ including leaves, $\eta_{p_j} = \eta \cdot 2^{-(\mathcal{D}- d_j)}$. $\eta$ for encoder $e$ and add-on layer $a$ was calculated by $\eta\cdot2^{-\mathcal{D}}$. Base $\eta$ for training was $\eta \in \{0.01, 0.005, 0.001\}$. Such sophisticated learning schedule is caused by the exponential increase of the computational complexity of gradients for the higher prototypes. It is common practice for Soft Decision Tree \cite{frosst2017distilling}. Our experiments were performed with NVIDIA RTX 2080 Ti.

\begin{table*}[ht]
\small
\caption{Results of molecular property prediction. Notice that ProGReST achieves better results than the baseline model (GCN) for each of the datasets. On $4$ out of $5$ datasets our model achieves the highest effectiveness. We conclude that interpretable molecular prediction can be done without a sacrifice of model performance.}
\begin{center}

\begin{tabular}{l c c c c c }
\hline
 $\text{Method}$
 & $\textbf{Caco-2} \downarrow$
 & $\textbf{PPBR} \downarrow$
 & $\textbf{LD50} \downarrow$
 & $\textbf{VDss} \uparrow$
 & $\textbf{HL} \uparrow$
 \\
\hline
RDKit2D + MLP~\cite{btaa1005} & $0.393 \pm 0.024$ & $9.994 \pm 0.319$ & $0.678 \pm 0.003$ & $0.561 \pm 0.025$ & $0.184 \pm 0.111$ \\
AttrMasking~\cite{weihua2019} & $0.546 \pm 0.052$ & $10.075 \pm 0.202$ & $0.685 \pm 0.025$ & $0.559 \pm 0.019 $ & $0.151 \pm 0.068 $ \\
Morgan + MLP~\cite{btaa1005} & $\text{--}$ & $12.848 \pm 0.362 $ & $0.649 \pm 0.019 $ & $0.493 \pm 0.011  $ & $0.329 \pm 0.083 $ \\
ContextPred~\cite{weihua2019} & $0.502 \pm 0.036 $ & $9.445 \pm 0.224 $ & $0.669 \pm 0.030 $ & $0.485 \pm 0.092 $ & $0.129 \pm 0.114$ \\
NeuralFP~\cite{lee2020} & $0.530 \pm 0.102 $ & $ 9.292 \pm 0.384 $ & $0.667 \pm 0.020  $ & $0.258 \pm 0.162 $ & $0.177 \pm 0.165 $ \\
AttentiveFP~\cite{Xiong20202} & $0.401 \pm 0.032$ & $9.373 \pm 0.335 $ & $0.678 \pm 0.012 $ & $0.241 \pm 0.145 $ & $0.085 \pm 0.068$ \\
CNN~\cite{btaa1005} & $0.446 \pm 0.036 $ & $11.106 \pm 0.358 $ & $0.675 \pm 0.011 $ & $0.226 \pm 0.114 $ & $0.038 \pm 0.138 $ \\
SimGCN~\cite{suman2020simgcn} & $\text{--}$ & $\text{--}$ & $\text{--}$ & $0.582 \pm 0.031$ & $ \mathbf{ 0.392 \pm 0.065} $ \\
\hline
ProGReST+GCN (Our) &  $0.367\pm0.022 $& $9.722\pm0.200 $ & $0.611\pm0.009 $ & $0.586\pm0.012 $ & $0.295\pm 0.058 $\\
GCN (Baseline) ~\cite{KipfW16} & $0.599 \pm 0.104$ & $10.194 \pm 0.373 $ & $0.649 \pm 0.026 $ & $0.457 \pm 0.050 $ & $0.239 \pm 0.100 $\\
\hline
ProGReST+RMAT (Our) &$\mathbf{ 0.360\pm0.069}$ & $\mathbf{ 9.256\pm 0.287} $ & $0.597 \pm0.072 $& $\mathbf{ 0.620 \pm 0.069} $ & $0.337\pm 0.049 $\\
RMAT~\cite{maziarka2021relative} & $0.363 \pm 0.030 $&$ 9.909 \pm 0.388$ & $\mathbf{  0.569 \pm 0.092 }$ & $0.487 \pm 0.083 $ & $0.360 \pm 0.063 $ \\

\hline
\end{tabular}
\label{tab_results}
\end{center}
\end{table*}
\section{Experiments}
We evaluate ProGReST on five datasets from the PyTDC repository~\cite{Huang2021tdc}: Caco-2~\cite{caco2}, PPBR~\cite{ppbr}, LD50~\cite{ld50}, VDss~\cite{vdss}, and Half-Life (HL)~\cite{Obach1385}. In the Supplementary Materials, we provide a short characteristic of the datasets. 

\paragraph{Results}

As Tab.~\ref{tab_results} shows, our ProGReST model not only outperforms its baseline model (GCN) but also achieves the best results on four of the five datasets. The prototypical-part-based approach for molecular activity prediction not only brings the interpretability of the predictions into the process, but also achieves superior results. This is in contrast to other prototypical-part-based methods such as~\cite{nauta2021neural} in computer vision where the introduction of interpretability reduces the model accuracy. 
\begin{figure}[t]
    \centering
    \includegraphics{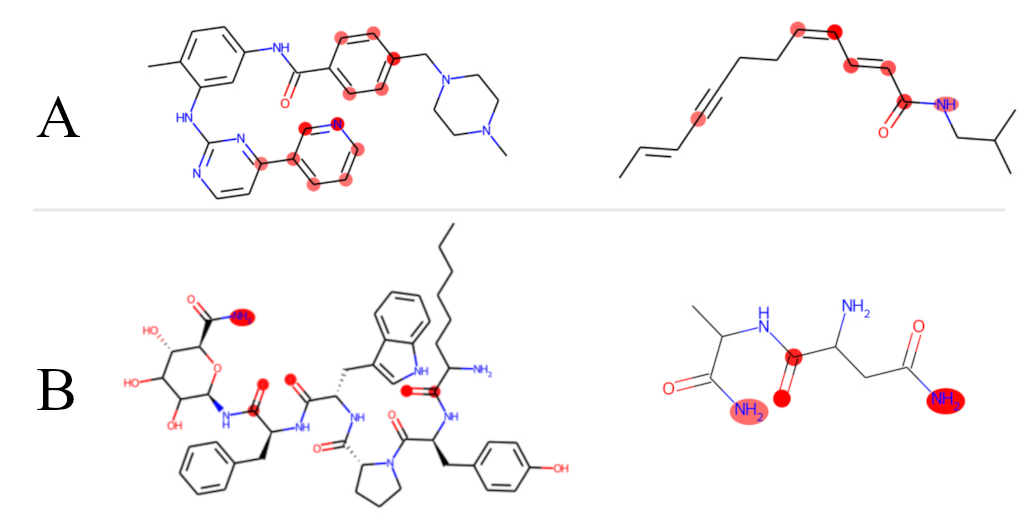}
    \caption{Two examples of the learned prototypical parts. The atoms that are marked with a red circle are a part of the prototype. The compounds on the left are the reference compounds, and the ones on the right are matched by the similarity of the prototypical parts. In prototype A, we see aromatic rings and conjugated bonds (alternating single and double bonds). Prototype B consists of ketones (=O) and amides (-C(=O)NH$_2$).}
    \label{fig:interpretation}
\end{figure}

\section{Interpretability}

To corroborate the interpretability of the prototypical parts learned by our model, we performed a qualitative study in which a chemist assessed the usefulness of the discovered molecular features. First, a small subset of compounds from the Caco-2 dataset was presented to the human expert, and the atoms constituting the learned prototypical parts were highlighted. Next, the compounds that contained the same prototypes were shown in order to confirm the agreement between the similarity function used in the model and the human knowledge-based intuition.

\begin{figure*}[h]
\centering
\begin{subfigure}{0.32\textwidth}
    \includegraphics[width=\textwidth]{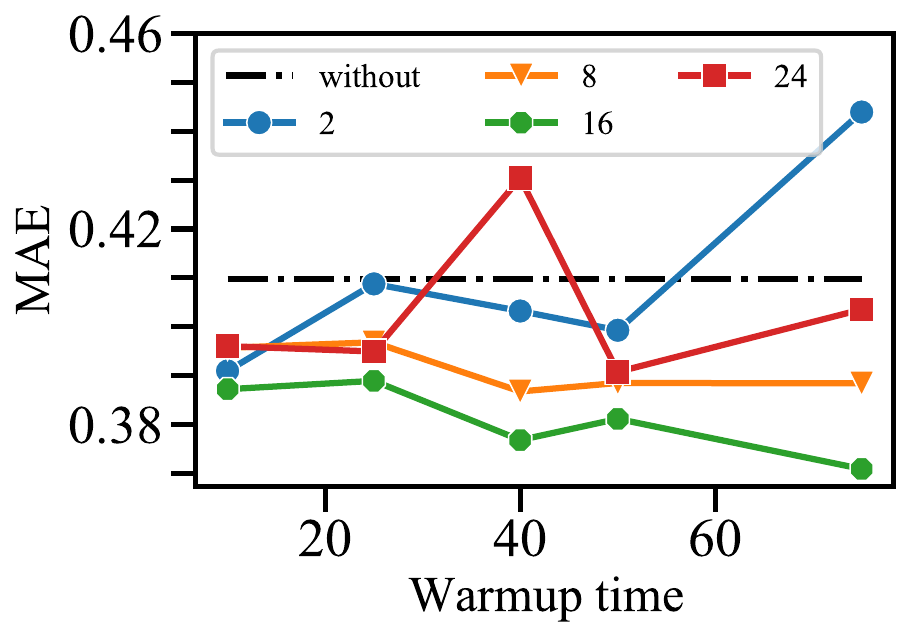}
    \caption{Effect of warming up for the Caco datase. Lower is better}
    \label{fig:warmup_caco}
\end{subfigure}
\hfill
\begin{subfigure}{0.32\textwidth}
    \includegraphics[width=\textwidth]{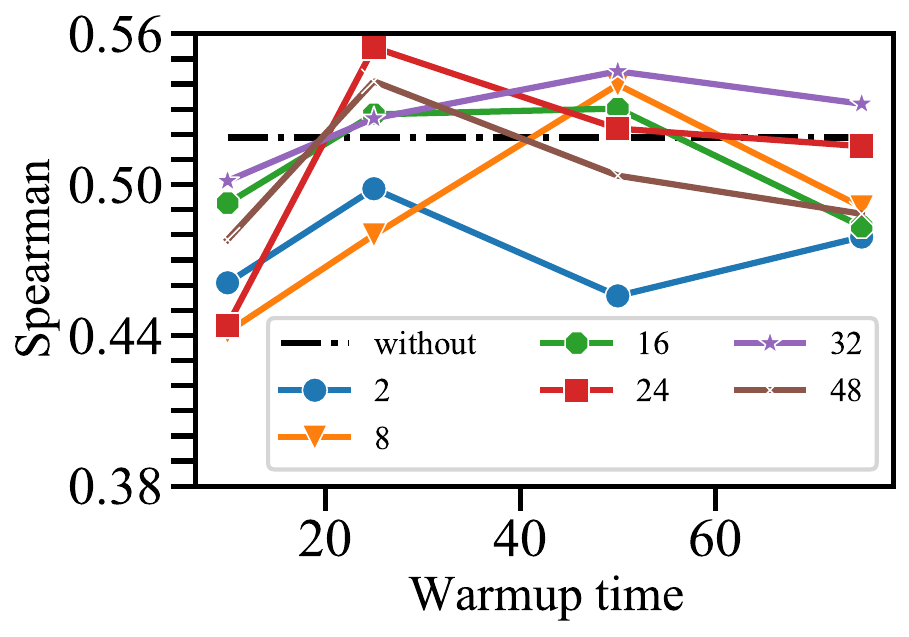}
    \caption{Effect of warming up for the VDss dataset. Higher is better.}
    \label{fig:warmup_vdss}
\end{subfigure}
\hfill
\begin{subfigure}{0.32\textwidth}
    \includegraphics[width=\textwidth]{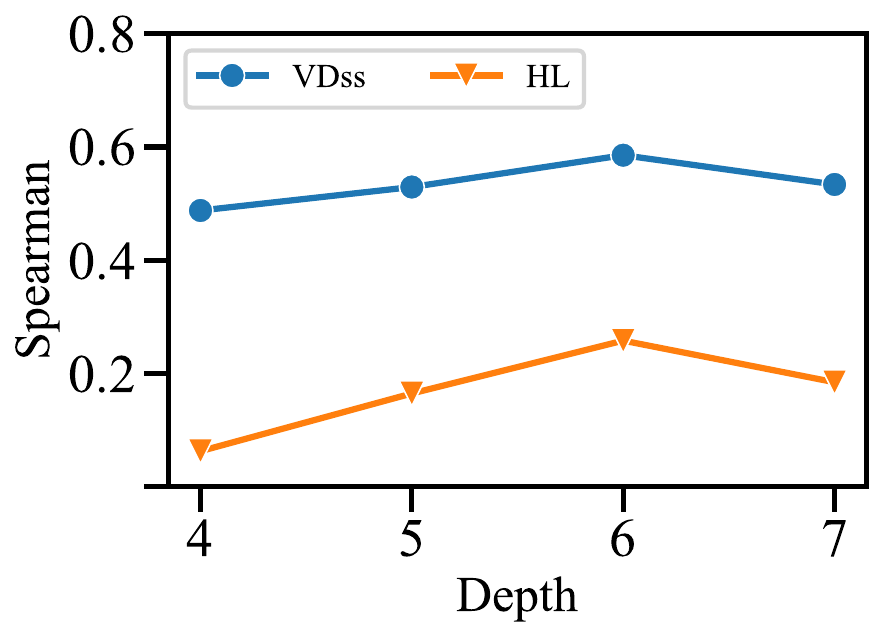}
    \caption{Effect of tree depth on model performance. Higher is better.}
    \label{fig:depth}
\end{subfigure}
        
\caption{The plots a) and b) shows how the number of clusters in K-Means in a warmup phase influences the model accuracy. We observe that, $K=24$ is the most optimal for VDss and $K=16$ is the most optimal for Caco. The plot c) show effect of tree depth on model performance.  We observe that the model performance saturates when the depth of the tree is increased over $6$. }
\label{fig:figures}
\end{figure*}

As a result of the visual inspection, a number of useful prototypical parts were identified. Caco-2 is a permeability assay, and there are several prominent molecular features that correlate well with the compound ability to penetrate the epithelial barrier. One of the features detected by the model is a set of ketone and amine groups that impact the hydrophilicity of the compound and can form hydrogen bonds with the lipid layers, which may hugely alter the permeability. In other prototypical parts we see aromatic rings or aliphatic side chains which are also related to the hydrophobicity and can be correlated with the compound bulkiness, decreasing the compound ability to pass the barrier. The described structures are depicted in Fig.~\ref{fig:interpretation}. 

The examination of the prototype similarity between different compounds showed that the same functional groups are correctly matched. What is more interesting, some more subtle similarities are also discovered by the model, e.g. conjugated bonds are found similar to the aromatic rings, indicating that the model captures the electronic nature of these structures. On the right side of Fig.~\ref{fig:interpretation}, an exemplary matching structure is shown.

\section{Ablation study}

\subsection{Warmup}
In this part, we present the influence of a warmup training stage on the effectiveness of our model. For two of the datasets (Caco2 and VDss), we show how the number of epochs and number of pseudo labels in the warmup phase influence the results, shown in Fig.~\ref{fig:warmup_vdss} and Fid.~\ref{fig:warmup_caco}. The results show that too long training is not beneficial for a warmup because the model overfits to labels derived from clustering and forgets features related to a regression task. Also, the number of pseudo-labels varies between datasets, for Caco2 the best one is $16$ while for VDss the best is $24$. It shows that these parameters should be chosen carefully, most probably due to the relatively small number of examples in each of the datasets. 

\begin{table}[t]
\caption{Comparison of the time needed to perform the projection. We conclude that Proxy Projection is much faster than MCTS-based projection.}
\begin{center}
\begin{tabular}{l c c}

\hline
 Depth & \textbf{4}
 & \textbf{5} \\
\hline
Proxy Projection & $\mathbf{10.1s}$ & $\mathbf{16.2s}$\\
MCTS-based Projection & $1760.0s$ & $4371.5s$  \\ 
\hline

\end{tabular}
\label{tab1_mcts}
\end{center}
\end{table}

\subsection{Proxy Projection vs MCTS-based projection}

In this part of ablation, we want to show a difference in computational time between Proxy and MCTS-based projections. As an MCTS-based one, we take the implementation from~\cite{zhang2021protgnn}. In contrast to ProtGNN, ProGReST can find the important vector of graph latent representation using a similarity function, because we do not perform pooling of a latent representation. Based on Tab.~\ref{tab1_mcts}, we conclude that MCTS-based projection works much slower than our novel Proxy Projection. That is why we recommend using it in periodical projections. To preserve the interpretability component, we encourage to use MCTS projection at the end of the training. 

\subsection{Depth of the tree}

Also, we checked how the depth of a tree influences the model performance. Too small trees don't have enough capacity to correctly model the task. While, too big ones tend to overfit and lost the ability to generalize well. We observed it for each dataset, as it is visible in Fig.~\ref{fig:depth}.

\subsection{Latent distance}

We checked the latent distance loss proposed in ProtGNN \cite{zhang2021protgnn}. The authors penalize the model if the distance between all prototypes assigned to a single data class is too small. However, in a regression task there are no classes. We decided to try to apply the proposed loss on all prototypes and it is inconclusive if it helps or not. However, the influence of the loss correlates with the depth of tree. For smaller trees it is beneficial to use this regularizers (for example Caco2 in Tab.~\ref{tab:parameters} has $5$ levels), while for deeper trees, the number of nodes grows exponentially. This makes it difficult to ensure orthogonality between all prototypes (for example LD50 in Tab.~\ref{tab:parameters} has $6$ levels and the score was worse than the one without the extra loss). That is why we used this loss only for parents of leaves in our tree. We can argue that some structures can exist independently of each other. For example the absence of an aromatic ring does not ensure the absence of any functional structure. 

\subsection{Influence of regularizers}

\begin{table*}[t]
\small
\caption{Influence of the different loss components on the model performance. Notice that the regularizers are crucial to training the model. A combination of all regularizers results in the best-performing model.}
\begin{center}

\begin{tabular}{l c c c c c}

\hline
 Dataset
 & \textbf{no regularizer}
 & \textbf{with cluster cost} 
 & \textbf{with path cost} 
 & \textbf{with latent distance cost} 
 & \textbf{all} 
 \\ 
\hline
Caco2 $\downarrow$  & $0.565\pm0.003  $ & $0.558\pm0.013 $ & $0.454\pm0.086 $ & $0.432\pm0.034$ &$ \mathbf{ 0.367\pm0.022} $\\
LD50 $\downarrow$  & $0.660\pm0.025  $ & $0.675\pm0.004 $ & $0.639\pm0.016 $ & $0.643\pm0.014$ &$ \mathbf{ 0.611\pm0.009} $\\
PPBR $\downarrow$ & $11.628\pm0.291$  & $11.272\pm0.701 $& $11.092\pm0.771 $& $9.846\pm0.407$ & $\mathbf{ 9.722\pm 0.200}$ \\

VDss $\uparrow$ & $0.130\pm0.125  $ & $0.180\pm0.231 $ & $0.180\pm0.151 $ & $0.531\pm0.064$ & $\mathbf{ 0.586\pm  0.012}$ \\

HL $\uparrow$ & $-0.031\pm0.051  $ & $0.033\pm0.134 $ & $0.029\pm0.098 $ & $0.193\pm0.099$ & $\mathbf{ 0.295\pm 0.058}$\\
\hline
\end{tabular}
\label{tab1_losses}
\end{center}
\end{table*}

As the training schema of the model is complicated, we tested the influence of each part of the loss function on the final performance of the model. The results are shown in Tab.~\ref{tab1_losses}. We notice that cluster cost does not increase the model performance, but it requires less epochs to train. For tested datasets, Caco2 with cluster cost needed $114\pm8$ epochs to get best result, but without it  $137\pm10$ epochs, LD50 with needed $184\pm28$, but without $215\pm19$. On the other hand, path cost with $\lambda_p$ greater than zero improves the model by encouraging the model to use all leaves during the training. The latent distance cost also results in better effectiveness of the model. We conclude that all of those regularizers are complementary and result in a lower error rate of the model.

\begin{table}[t]
\small
\caption{Different strategies for distance loss. One can observe, that regularizing only parent nodes is the most effective to obtain the best model.}
\label{tab:parameters}
\begin{center}

\begin{tabular}{l c c c}

\hline
 Dataset
 & \textbf{without}
 & \textbf{parents} 
 & \textbf{all\cite{zhang2021protgnn}}
 \\
\hline
Caco2 $\downarrow$  & $0.402 \pm 0.032 $&$ \mathbf{ 0.367 \pm 0.022} $ & $0.387 \pm 0.030$\\
LD50 $\downarrow$ & $0.625 \pm 0.228 $& $\mathbf{ 0.611 \pm 0.009}  $ & $0.632 \pm 0.025$\\
VDss $\uparrow$ & $0.544 \pm 0.051$ & $\mathbf{ 0.586 \pm 0.012}  $ & $0.558 \pm 0.044$\\
HL $\uparrow$ & $0.239 \pm 0.025 $&$ \mathbf{ 0.295 \pm 0.058}  $ & $0.212 \pm 0.053$\\
\hline
\end{tabular}
\label{tab1}
\end{center}
\end{table}

\subsection{Pretrained model}

As most prototype-based models use pretrained models in their backbones, especially those in the computer vision domain. We investigated how the usage of a pretrained model on a large chemical database behaves in a prototype-based learning scenario. We decided to use R-MAT~\cite{maziarka2020molecule}, the current state-of-the-art model for chemical compound representation. We observe the improvement of the results for all datasets, as Tab.~\ref{tab_results} shows. 

\section{Conclusions}
In this work, we introduce ProGReST, which is an interpretable model for regression of molecular properties. We show that it not only brings the interpretability into the prediction but also outperforms GCN architecture, as well as achieves state-of-the-art results on $4$ out of $5$ datasets that we tested. Additionally, we introduce a proxy projection which accelerates the training time and reduces the power consumption needed for training which is important from an environmental point of view. Finally, we show that the ProGReST explanations are valid and show the influence of the novelties and their hyperparameters on the models' performance.

 In future works, we want to analyze the possibility of pruning for ProGReST and generalize it to other data types such as text, because in NLP there are a lot of problems represented as graphs and regression tasks.
 
\section*{Acknowledgments}
 
The work of D. Rymarczyk and D. Dobrowolski is supported by the National Centre of Science (Poland) Grant No. 2021/41/B/ST6/01370, and the work of T. Danel is supported by National Centre of Science (Poland) Grant No. 2020/37/N/ST6/02728.

\bibliographystyle{siam}

\bibliography{dsaa}

\newpage

\section*{Algorithm}

\begin{algorithm}[h!]
\caption{Training a ProGReST}
\begin{algorithmic}[1]
\Require  Dataset, depth $d$, $nEpoch$, $nWarmup$, $c$ classes, $\lambda_c$, $\lambda_d$, $\lambda_p$, projection start $T_s$ and how often to perform $T_m$

\State Initialize ProGReST model with depth $d$ and trainable parameters $w^l$ for leaves, $w_f$, $p$ for the rest of the model 
\State Create $\mathcal{K}$ K-Means for warmup with $c$ classes from train dataset

\For{$e \in \{1, ..., nWarmup$\}} \Comment{Warmup} 
    \For{\textbf{each} batch in train dataset}
        \State calculate $\mathcal{L} _{CE}$ for current batch
       \State and
       update $w^l$, $w_f$, $p$  with gradient descent
    \EndFor
\EndFor
\State translate each $l \in \ell$ to scalar using Eq. \ref{eq:translate_leaf} formula 
\For{$e \in \{nWarmup, ..., nEpochs$\}} \Comment{Training} 

\For{\textbf{each} batch in train dataset}
    \If{$e \geq T_s$ and $e \% T_m = 0$}
            \State project prototypes using Eq. \ref{eq:project_patch}
      \EndIf
    \State calculate $\mathcal{L}$ for current batch
    \State and update $w^l$, $w_f$, $p$  with gradient descent
\EndFor
\EndFor
~project prototypes using Eq. \ref{eq:project_mcts}
 \For{\textbf{each} batch in train dataset}
        \State calculate $\mathcal{L}$ for current batch
       \State and update $w^l$  with gradient descent
\EndFor
\end{algorithmic}
\end{algorithm}

\section*{Datasets}

\paragraph{Caco-2: Cell Effective Permeability \cite{caco2}} 

Caco-2 is a human colon epithelial cancer cell line used as an in vitro model to simulate the human intestinal tissue. The experimental result on the rate of drug passing through the Caco-2 cells can approximate the rate at which the drug permeates through the human intestinal tissue. The dataset consists of $906$ drugs and is evaluated using the Mean Absolute Error metric, and employs a scaffold-based fold splitting strategy.

\paragraph{PPBR - Plasma Protein Binding Rate \cite{ppbr}} 

PPBR contains data about expression of a drug bound to plasma proteins in the blood. It strongly affects a drug's efficiency of delivery. The less bound a drug is, the more efficiently it can traverse and diffuse to the site of actions. It is derived from a ChEMBL~\cite{gaulton2017chembl} assay deposited by AstraZeneca. It contains 1,797 drugs and is evaluated using the Mean Absolute Error metric, and employs a scaffold-based fold splitting strategy.
  
\paragraph{LD50 - Acute Toxicity \cite{ld50}} 

Acute toxicity LD50 measures the most conservative dose that can lead to lethal adverse effects. The higher the dose, the more lethal a drug. It is a very important measure showing how much of a drug a patient can take to not have serious side effects. The dataset consists of 7,385 drugs, is evaluated using MAE, and employs a scaffold-based fold splitting strategy.

\paragraph{VDss  - Volume of Distribution at steady state \cite{vdss}} 

 The volume of distribution at steady state (VDss) measures the degree of a drug's concentration in body tissue compared to the concentration in blood. Higher VD indicates a higher distribution in the tissue and usually indicates the drug with high lipid solubility, and a low plasma protein binding rate. The dataset contains 1,130 drugs and uses Spearman correlation as an evaluation metric, and employs a scaffold-based fold splitting strategy.

\paragraph{Half Life (HL) \cite{Obach1385}} 

 The half-life of a drug is the duration for the concentration of the drug in the body to be reduced by half. It measures the duration of actions of a drug.  It contains 667 drugs and uses Spearman correlation as an evaluation metric, and employs a scaffold-based fold splitting strategy.

\end{document}